\documentclass[a4paper,11pt]{article}
\usepackage{pos}

\title{Slow diffusion around pulsar $\gamma$-ray halos and its impact on cosmic rays propagation}
 \ShortTitle{Short Title for header}

\author*{Xiao-Jun Bi}

\affiliation{Institute of High Energy Physics, Chinese Academy of Sciences\\
  Yuquan Road 19B, Beijing, P.R. China}

\emailAdd{bixj@ihep.ac.cn}

\abstract{The diffusion coefficients around the pulsar $\gamma$-ray halos are highly suppressed compared with the value in the interstellar medium. It is suggested in the literature that the $\gamma$-ray halos can be explained by a ballistic-diffusive (BD) propagation without slow diffusion. However our calculation shows that the BD propagation can not account for the $\gamma$-ray halo profile well. Furthermore the transfer efficiency of the pulsar spin down energy to the high energy electrons and positrons is even larger than 1 in the BD scenario. Therefore slow diffusion is necessary to account for the pulsar $\gamma$-ray halos. Taking the slow diffusion into account the contribution of positron flux originated from nearby pulsars to the AMS-02 data is reexamined. We may also expect a slow diffusion disk of the Milky Way as many such slow diffusion regions exist. The positron contribution to the AMS-02 data from dark matter annihilation in the new propagation model is also reexamined. We find that the dark matter  scenario satisfies all the $\gamma$-ray limits in the new propagation model.}

\ConferenceLogo{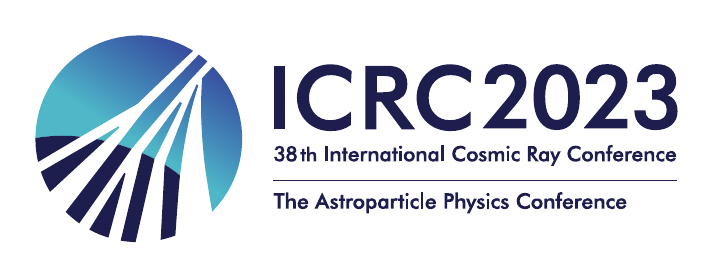}

\FullConference{%
38th International Cosmic Ray Conference (ICRC2023)\\
  26 July - 3 August, 2023\\
  Nagoya, Japan}

\begin{document}
\maketitle

\section{Introduction}

In 2017 the $\gamma$-ray halos around the Geminga and Monogem pulsar wind nebula (PWN) were first detected by the HAWC collaboration \cite{HAWC:2017kbo}. The $\gamma$-ray halos are generated by high energy electrons and positrons accelerated by the PWN scattering with the CMB when diffusing away from the central PWN. 
The surface brightness profile (SBP) of the halos were measured as function of distance from the PWN. According to the profile of the surface brightness the electron/positron distribution around the PWN was determined accordingly and the diffusion coefficient is derived then. 
The most important conclusion of the HAWC observation is that the diffusion coefficient is hundreds times smaller than the value at the interstellar medium which is determined by the cosmic ray B/C data \cite{Yuan:2017ozr}.

Recently the LHAASO collaboration observed another pulsar $\gamma$-ray halo around PSR J0622+3749 \cite{LHAASO:2021crt}. The diffusion coefficient around PSR J0622+3749 is similar to that around Geminga. The observation confirms the conjecture that the $\gamma$-ray halos may be a universal structure at the middle to old age of a PWN \cite{Fang:2019iym}.

However, if the slow diffusion is uniform in the Galaxy it will lead to disastrous consequences as all the predictions by the diffusive propagation model are wrong, as all the secondary products like B, Li, Be and $\bar{p}$, $e^+$ and diffuse $\gamma$-rays will be improved by orders  of magnitude. This is certainly unacceptable. Therefore a two-zone diffusion model is proposed \cite{Fang:2018qco}. In the two-zone model the slow diffusion regions is a small region around the pulsar with a radius $r_*$. Outside of the slow diffusion region the diffusion coefficient is the regular value. Therefore all the correct predictions of the secondary products are kept unchanged as the slow diffusion region is only around the pulsars. Especially it is found that the contribution of positron flux to the AMS-02 data from Geminga is enhanced in the two-zone diffusion model compared with the conventional fast diffusion model \cite{Fang:2018qco}. This is a happy surprise which has not been expected. The result makes the pulsar origin of positrons to account for the positron excess more feasible. 

It is a very important question that what mechanism leads to such a suppressed diffusion rate of charged particles. It was first suggested the suppression is due to the streaming instability generated by the electrons and positrons injected from the PWN to the surrounding medium \cite{Evoli:2018aza}. However, the self generated $\gamma$-ray halo is not powerful enough to suppressed the diffusion coefficient to the observed value. Later this conclusion was proved analytically \cite{Fang:2019iym}. In \cite{Fang:2019iym} it is suggested that Geminga is still within its precursor SNR. As the expansion of the SNR the shock waves interacts with the interstellar medium and stimulate the magnetic turbulence. The shocked region in the down stream has strong magnetic turbulence and therefore a suppressed diffusion coefficient for charged particles. This mechanism explains the observed diffusion coefficient well \cite{Fang:2019iym}.

Recently there is a very interesting suggestion in the literature that ballistic-diffusive propagation may account for the $\gamma$-ray halos without introducing highly  suppressed diffusion \cite{Recchia:2021kty}. We reexamined this scenario and found the BD propagation can not account for the $\gamma$-ray halos and slow diffusion is necessary to account for the pulsar $\gamma$-ray halos \cite{Bao:2021hey}.

The slow diffusion region around pulsars has important impact on the cosmic rays propagation. We considered two aspects of the impact. First we consider how the slow diffusion region around pulsars affect the contribution of positrons from nearly pulsars to the AMS-02 data and explain the positron excess \cite{Fang:2019ayz}. Secondly we consider a new cosmic rays propagation model with a slow disk \cite{Zhao:2021yzf} and reexamine the dark matter scenario accounting for the positron excess in the new propagation model \cite{Lv:2023alj}. We find the dark matter scenario works well in the new propagation model and satisfies all the constraints on the dark matter annihilation by $\gamma$-ray observations \cite{Liu:2016ngs}.


\section{Slow diffusion is necessary for pulsar $\gamma$-ray halos}

\begin{figure}
	\centering
	\includegraphics[width=10cm]{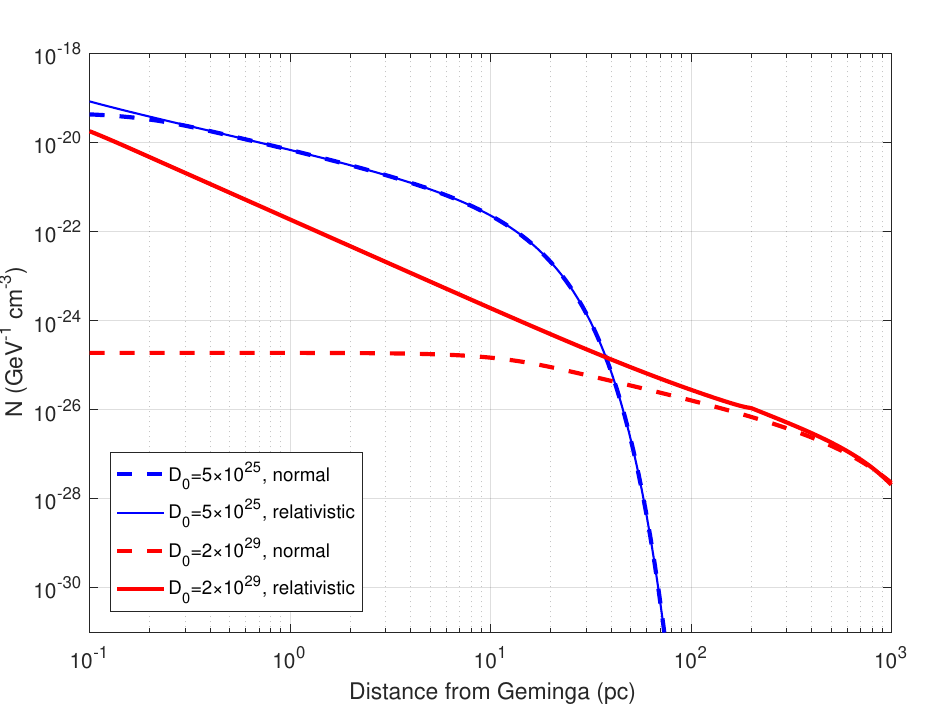}
	\caption{Electron number densities as functions of the distance from the central pulsar for the slow-diffusion ($D_0 = 5\times10^{25}$ cm$^2$ s$^{-1}$) and fast-diffusion ($D_0 = 2\times10^{29}$ cm$^2$ s$^{-1}$) scenarios. Both the relativistic and non-relativistic diffusion models are presented. The conversion efficiency is set as 100\% for all the cases in this figure. The figure is from Ref. \cite{Bao:2021hey}.
	\label{fig:density}}
\end{figure}

Recently Recchia et al. proposed an interesting mechanism to account for the observed pulsar $\gamma$-ray halos \cite{Recchia:2021kty} in the conventional fast diffusion propagation. The main idea of the mechanism is as following. As the diffusive propagation of cosmic rays (CRs) is described by a non-relativistic equation, the most recent injected particles for $t < 3D/c^2$ transport superluminally. Therefore the electron density around the pulsar is actually suppressed. Actually the most recently injected particles should propagate ballistically within the distance around the coherence length of the magnetic field. For distance larger than the coherence length the particles propagation is described by diffusion process. Taking the ballistic propagation into account and later transferring to the diffusion process the authors in \cite{Recchia:2021kty} point out the pulsar $\gamma$-ray halos may be accounted for in this scenario. The idea is shown in Fig. \ref{fig:density} clearly. The red dashed line shows the electron density distribution as function of distance to the pulsar in the diffusion process, while the solid red line shows the distribution in the ballistic-diffusive (BD) scenario. The blue lines show the distribution in the slow diffusion scenarios. It is clearly shown that for fast diffusion the BD scenario has a sharp peak at the center compared with that in pure fast diffusion scenario. For the slow diffusion the two scenarios have little difference. 

\begin{figure}
	\centering
	\includegraphics[width=7cm]{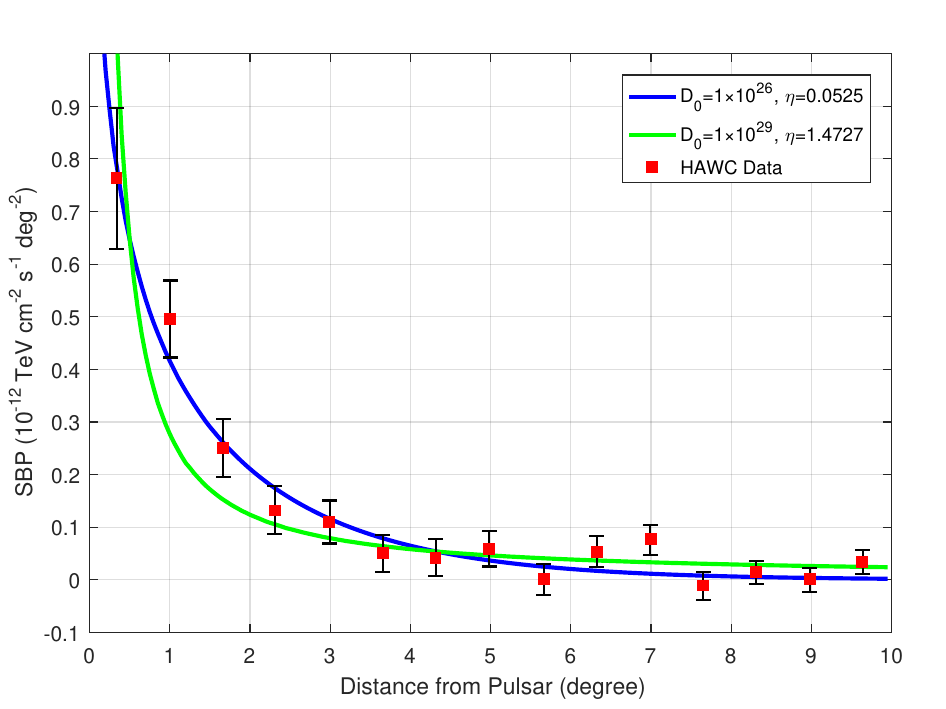}
    \includegraphics[width=7cm]{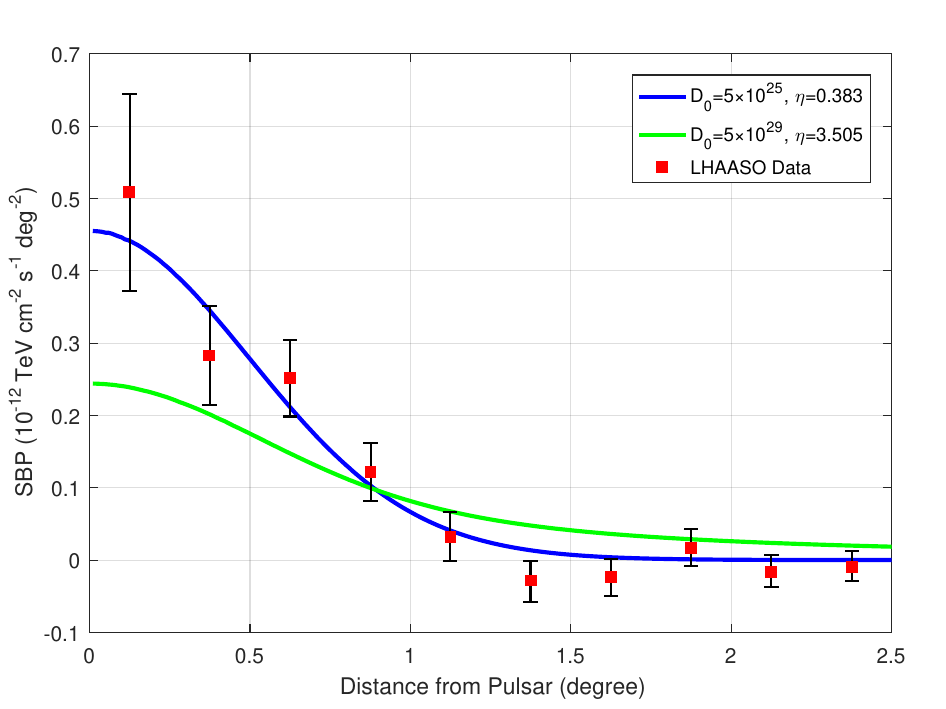}
	\caption{ The $\gamma$-ray profiles of the Geminga halo (left) and LHAASO J0621$+$3755 (right) taking $D_0$ corresponding to the two local minimal $\chi^2$. The figure is from Ref. \cite{Bao:2021hey}.}
	\label{fig:profile}
\end{figure}

We reexamined the BD scenario in \cite{Bao:2021hey}. In Fig. \ref{fig:profile} the surface brightness profile (SBP) of Geminga (left panel) and LHAASO J0621+3755 (right panel) are shown for both the BD (green line) and slow diffusion (blue) scenarios. It is easily seen that the BD scenario can not explain the HAWC data very well.

\begin{figure}
	\centering
	\includegraphics[width=7cm]{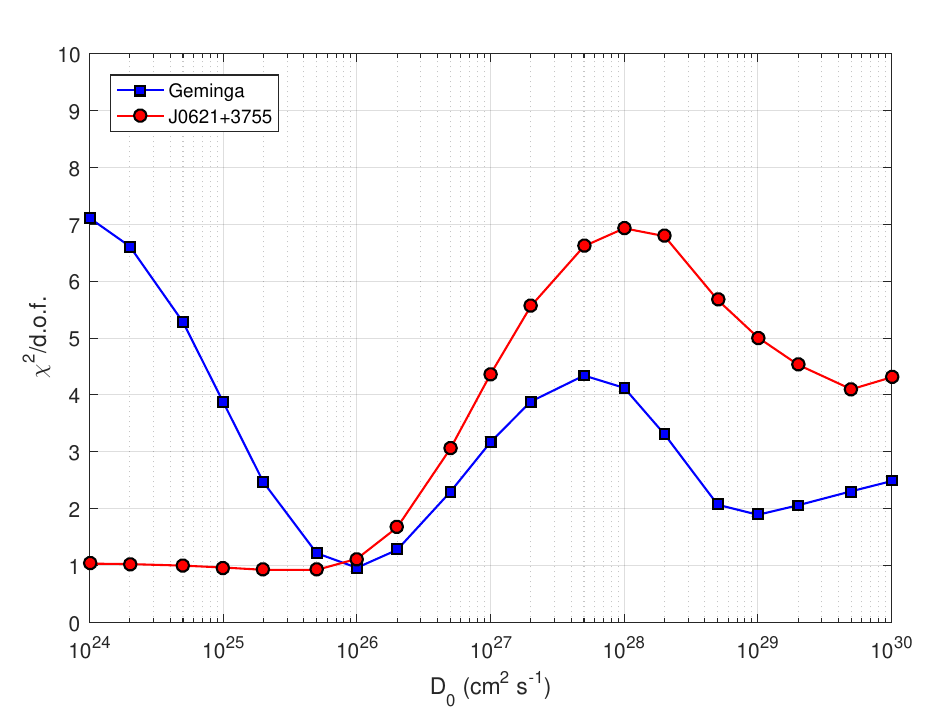}
	\includegraphics[width=7cm]{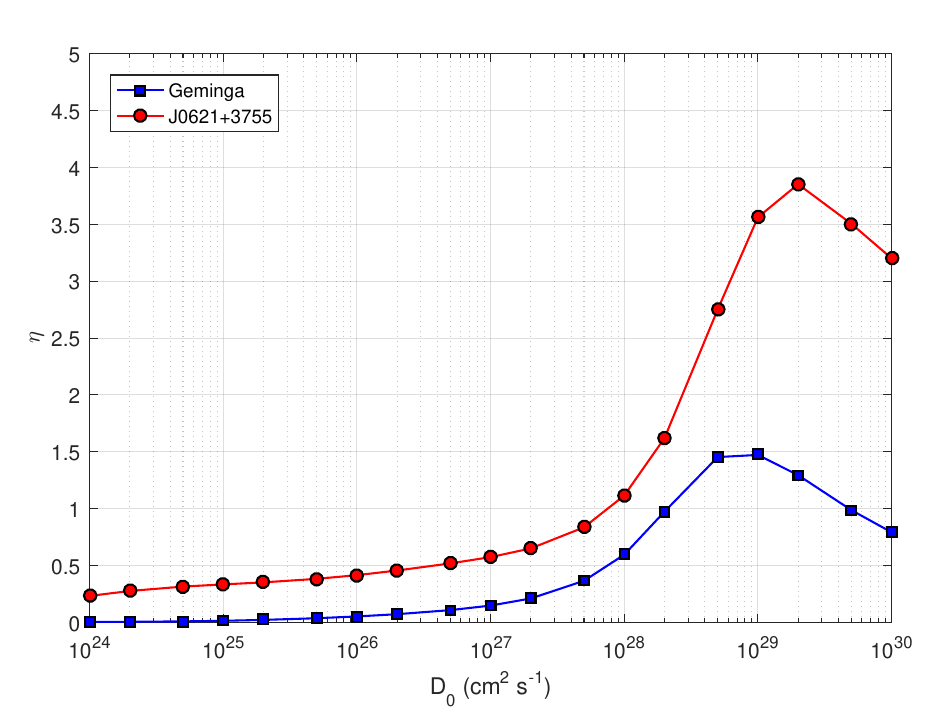}
	\caption{Left: The minimal $\chi^2$ of best fitting to the $\gamma$-ray halo profiles as function of diffusion coefficient at 1 GeV. Right: The efficiency of energy transfer from the pulsar spin-down energy to electron/positron pairs for different values of $D_0$. The figure is from Ref. \cite{Bao:2021hey}.}
	\label{fig:chi2}
\end{figure}

In the left panel of Fig. \ref{fig:chi2} the $\chi^2$ for the fitting of the HAWC data for different values of the diffusion coefficients are shown. The red curve is for LHAASO J0621+3755 and blue curve is for Geminga. The curves do show two minimal for small and large diffusion coefficients. For the slow diffusion the $\chi^2/d.o.f.$ is nearly equal to 1 while for large diffusion coefficient the minimal $\chi^2/d.o.f.$ are about  2 and 4 for Geminga and  LHAASO J0621+3755 respectively. Therefore it is shown the BD propagation can not account for the pulsar $\gamma$-ray profile.

\begin{figure}
	\centering
	\includegraphics[width=8.6cm]{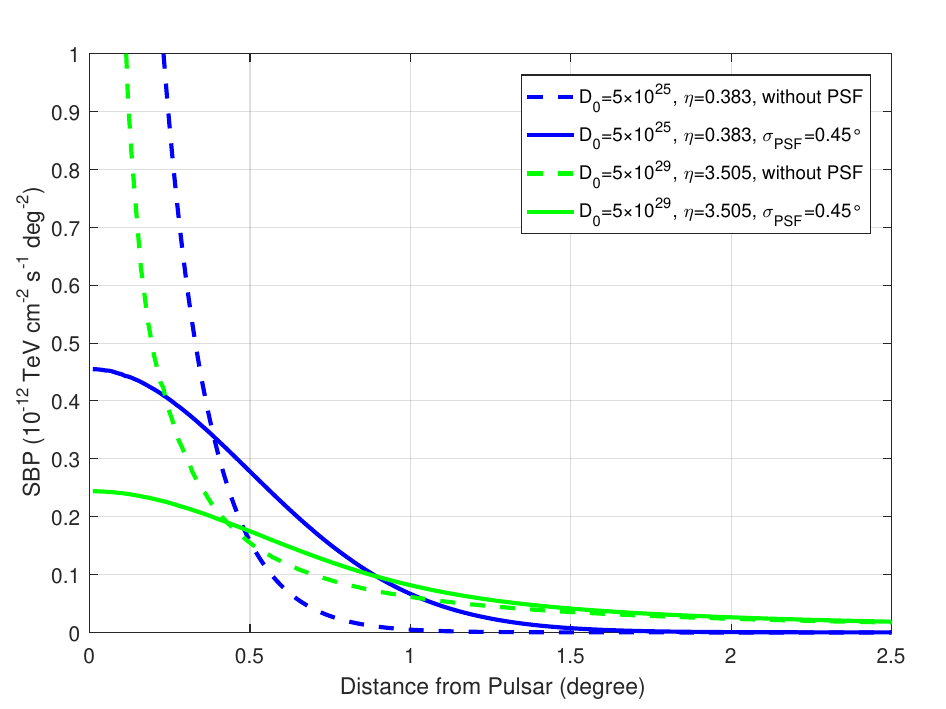}
	\caption{ Comparison of the $\gamma$-ray profiles for LHAASO J0621$+$3755 before and after PSF convolution. The figure is from Ref. \cite{Bao:2021hey}. }
	\label{fig:psf}
\end{figure}

The result can be understood from Fig. \ref{fig:psf}. The dashed lines in Fig. \ref{fig:psf} are the theoretical prediction of the SBP for slow and fast diffusion while the solid lines are the SBP after convolution with the PSF of LHAASO detector. It is clearly shown that the BD model predicts much flatter profile after convolving with PSF since it diffuses much faster than the slow diffusion case.

Further in the right panel of Fig. \ref{fig:chi2} we show the efficiency of the pulsar spin-down energy transferring to high energy electrons and positrons to account for the SBP of the $\gamma$-ray halos. The efficiencies for slow diffusion cases are reasonable while it is too large for the fast diffusion cases. Therefore we give conclusion that {\bf slow diffusion is necessary to account for the pulsar $\gamma$-ray halos}.

\section{Nearby pulsar to account for the positron excess}

In the HAWC paper \cite{HAWC:2017kbo} another important conclusion is that the positron flux contributed by Geminga in the slow diffusion scenario is negligible. This conclusion excludes the possibility that pulsars contribution accounts for the positron excess \cite{Yin:2013vaa}. This difficulty is overcame in the two-zone diffusion model \cite{Fang:2018qco} as described in Introduction. The existence of slow diffusion around a pulsar has important impact on cosmic rays propagation. In the following we discuss two aspects of the impact, that is, the impact on an individual source and impact on the Galactic cosmic rays propagation globally. 

\begin{figure}[t]
 \centering
 \includegraphics[width=0.6\textwidth]{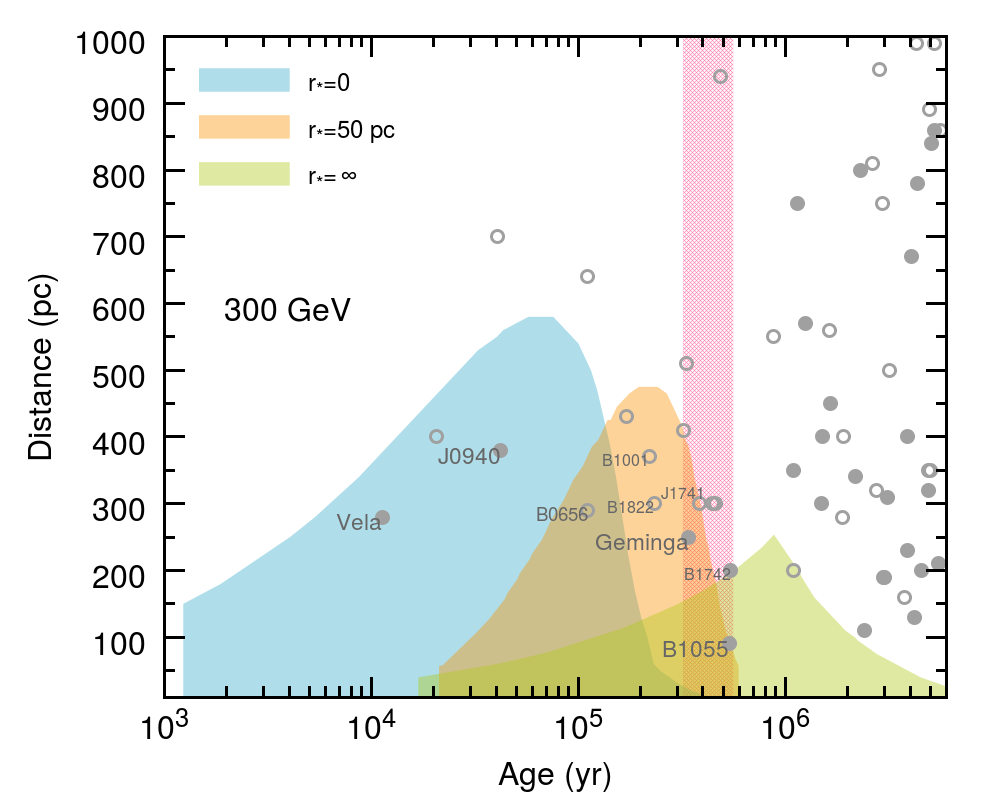}
 \caption{Pulsar distribution as a function of age and 
distance to the Earth. The shaded regions represent 
the pulsars that can contribute more than half of the measured positron flux at 300 GeV, and different colors represent different diffusion 
models. The red band shows the age range corresponding to positron spectral cut. Figure from \cite{Fang:2019ayz}. }
 \label{fig:contour_300gev}
\end{figure}

\begin{figure}[t]
 \centering
 \includegraphics[width=0.48\textwidth]{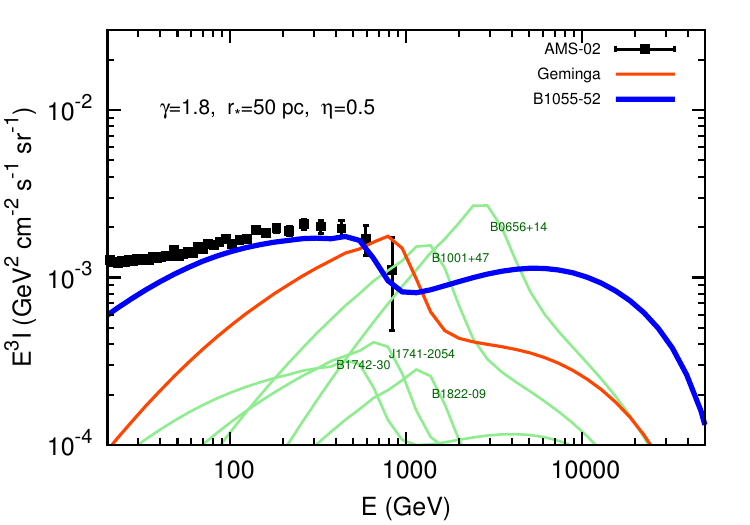}
 \caption{Positron spectra from the nearby and middle-aged pulsars under 
the two-zone diffusion model, compared with the latest AMS-02 data. Figure from \cite{Fang:2019ayz}.}
 \label{fig:2zone_50_1d8}
\end{figure}

In Ref. \cite{Fang:2019ayz}  we reanalyzed the possible pulsar that can account for the positron excess in the two-zone diffusion model. In Fig. \ref{fig:contour_300gev} we show the pulsars distribution with age and distance to the Earth. The red band is the age range in which the cooling effect leads to a cutoff at the electron/positron spectrum  consistent with the observed positron spectrum cutoff by the latest AMS-02 measurement. The shaded regions show the proper parameters that is possible to account for the positron flux for the conventional  case ($r_*=0$),  slow diffusion case ($r_*= \infty $) and for the two-zone diffusion case with $r_*=50$pc.

By checking the pulsars whose parameters are in the overlap region between the red band and the shaded region of two-zone diffusion we find an idea candidate pulsar B1055-52. In Fig. \ref{fig:2zone_50_1d8}  we show the positron spectrum from different pulsars. We notice the spectrum from the pulsar  B1055-52 with age about 500,000 years has a cutoff perfectly fit the AMS-02 data. The spectrum from Geminga with age about 300,000 years has a slightly higher cutoff. Anyway, the calculation shows that {\bf in the two-zone diffusion model an older  and closer pulsar, like B1055-52 ($t\sim 500$kyr, $d\sim 90$pc), is an ideal sources for the positron excess}. 

{\bf Note:}  During my presentation an audience gave a comment. He claimed the distance of pulsar B1055-52 was modified recently. However, we did not find such literature after the ICRC. A most recent paper claims the distance of B1055 is uncertain and 90pc is still a possible value \cite{Posselt:2023dau}. Even the distance of B1055 is determined different from 90pc in the future, this does not change our conclusion. In Fig. \ref{fig:contour_300gev} we show the correct parameter region for pulsars which can account for the AMS-02 positron spectrum. Even B1055-52 is not fell in this region another pulsar in the region can make similar contribution. 

\section{Dark matter annihilation for positron excess in the slow diffusion disk propagation model}

After the discovery of HAWC, there were many works in the literary about the $\gamma$-ray halos. It is now believed that the $\gamma$-ray halos are generally formed in the middle and old aged pulsars. Considering that 1-3 supernovas are generated each 100 years in the Milky Way with a pulsar lifetime $\sim 10^6$ years, there are about $\sim 10^4$ such slow diffusion regions with $r_*\sim 50$pc, which take significant fraction at the Galactic disk. Therefore we may expect the disk diffusion on average is slower than that outside of the disk. 

In Ref.\cite{Zhao:2021yzf}  we studied this cosmic rays propagation model with a slow disk. By fitting the AMS-02 data of C flux, B/C and Be/B by the MCMC simulation, the model parameters are determined precisely. It is very interesting that the diffusion in the disk is suppressed by fitting to data without introducing a slow disk in advance. Many features in the spectra of AMS-02 data are naturally explained. Especially the cosmic rays anisotropy are suppressed greatly in the slow disk model and is consistent with observations \cite{Zhao:2021yzf}. 

We reexamined the dark matter model to explain the positron excess in this slow disk propagation model \cite{Lv:2023alj}. The dark matter scenario has been studied extensively after the discovery of positron excess by PAMELA and AMS-02. However, previous studies show that the DM scenario is disfavored strongly by the $\gamma$-ray observations by Fermi-LAT and by the CMB observation by Planck \cite{Xiang:2017jou}. 

\begin{figure}[t]
 \centering
 \includegraphics[width=0.48\textwidth]{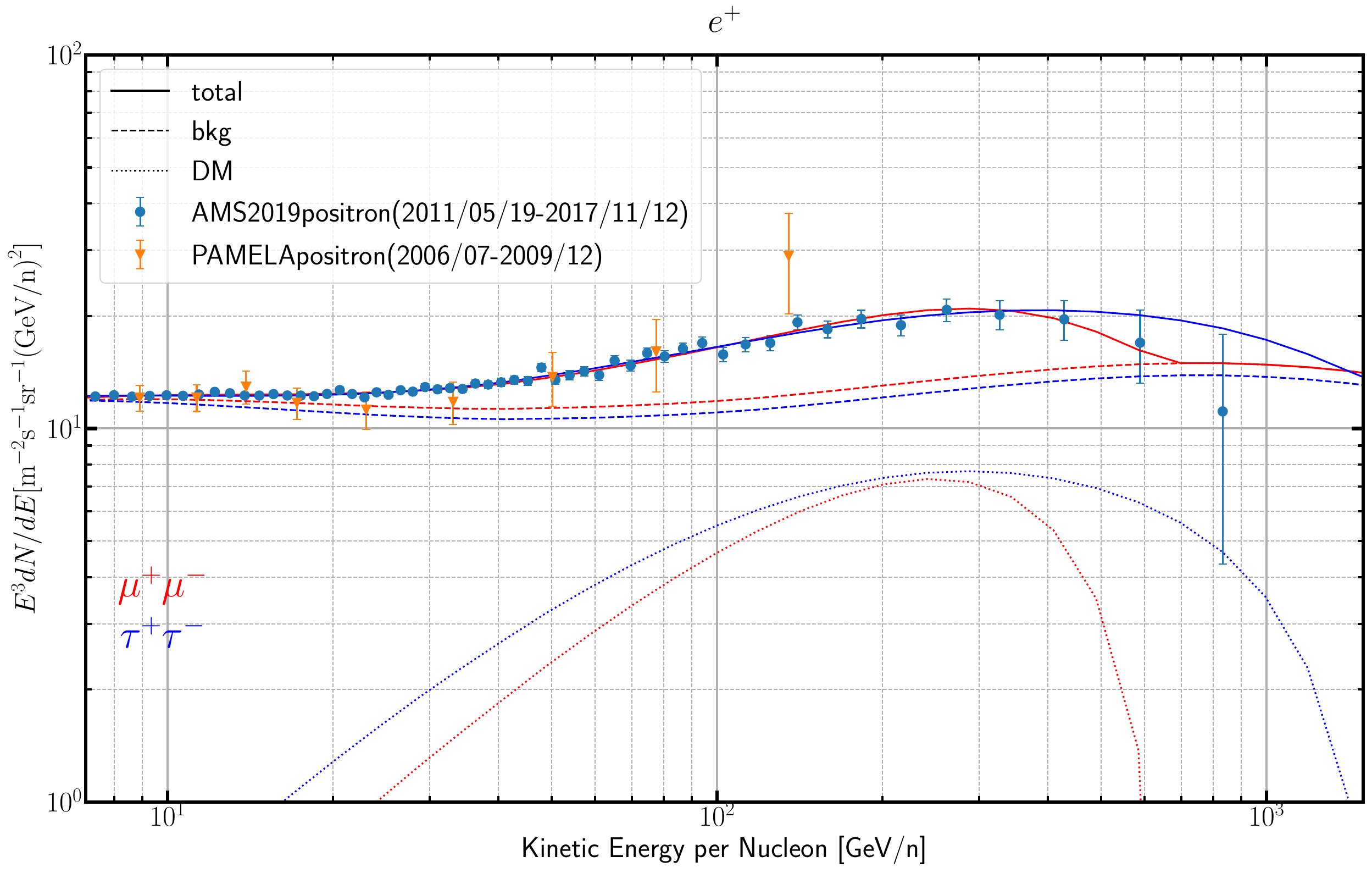}
 \caption{ The positron spectra taking DM annihilation into account with the best-fit parameters. AMS-02 data are shown. 
     The dashed, dotted, and solid lines represent the backgrounds, DM contributions, and total results, respectively. Figure from \cite{Lv:2023alj}. }
    \label{fig:ann}
 \end{figure}

\begin{figure}[t]
 \centering
 \includegraphics[width=0.48\textwidth]{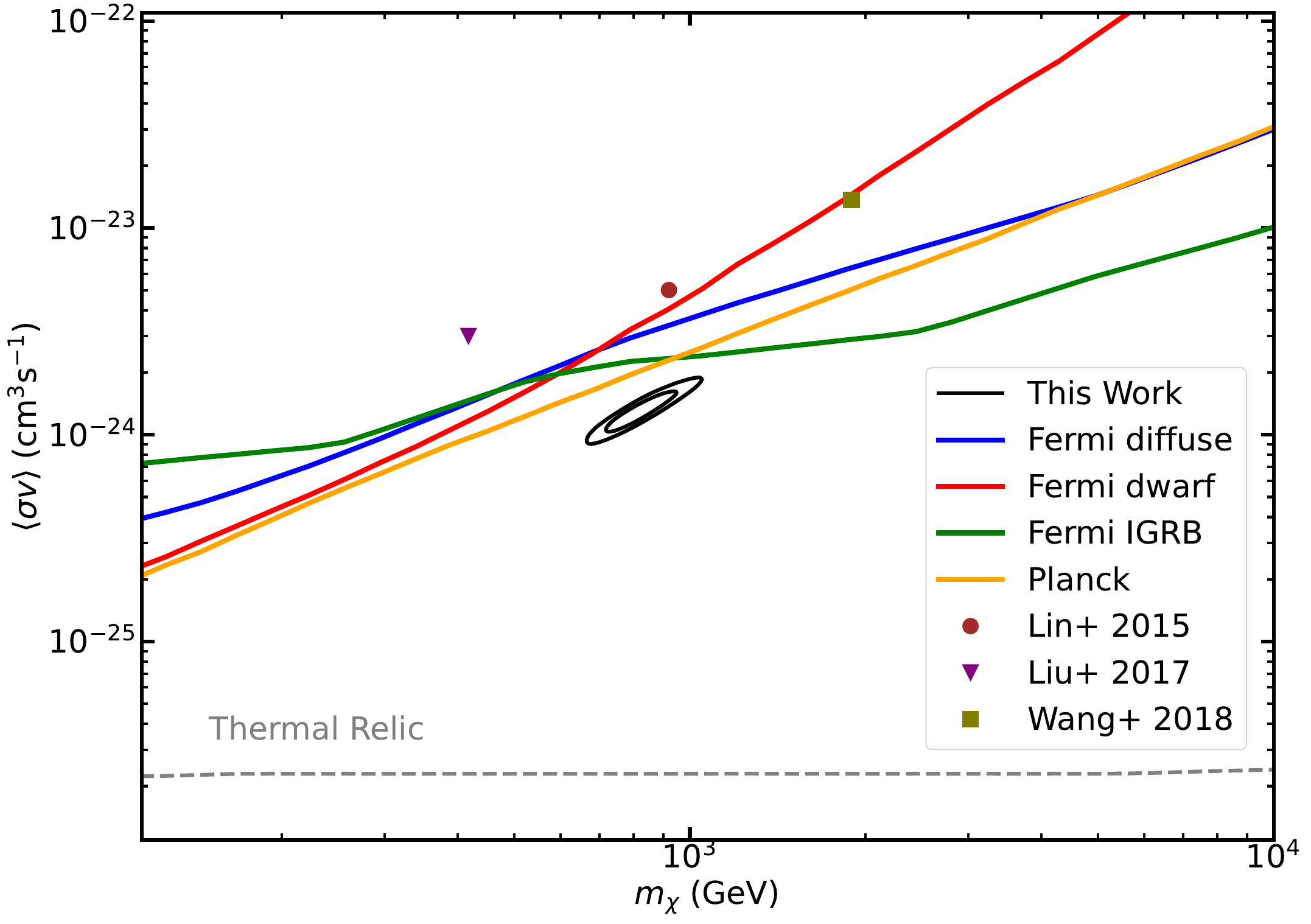}
 \caption{ $1\sigma$ and $2\sigma$ confidence regions in the $m_{\mathrm{DM}}-\langle\sigma v\rangle$ plane to fit the AMS-02 data, together with the exclusion lines from the Fermi observations of dwarf galaxies~\cite{Fermi-LAT:2016uux}, diffuse gamma-rays in the Milky Way halo~\cite{Fermi-LAT:2012pls}, IGRB~\cite{Liu:2016ngs}, and the Planck CMB observations~\cite{Slatyer:2015jla}. The fitting results to the AMS-02 observations from some previous analyses~\cite{Lin:2014vja,Liu:2016ngs,Wang:2018pcc} are shown as colored points. Figure from \cite{Lv:2023alj}. }
 \label{fig:dm}
 \end{figure}

In Fig. \ref{fig:ann}  we show the positron spectrum from DM annihilation into $\mu^+\mu^-$ and $\tau^+\tau^-$ final states taking the best fit parameters. In Fig. \ref{fig:dm} the contour is the parameter region that fits the AMS-02 positrons spectrum. The best fit values are $m_\chi = 770 $GeV and $\langle \sigma v\rangle = 9.3\times 10^{-25} cm^3/s$ for the dark matter mass and annihilation cross section respectively. In Fig. \ref{fig:dm} it is shown clearly that {\bf in the new propagation model the favored region of DM annihilation satisfies all the constraints from $\gamma$-ray observations}.  

The reason for smaller annihilation cross section in the slow disk propagation model is easy to understand. Because of the slow disk the secondary positrons and that from dark matter annihilation are more confined in the disk. Therefore a smaller annihilation cross section can account for the positron flux measured by AMS-02.

\section{Summary}

In summary, we find that a slow diffusion region around the pulsar is necessary to account for the surrounding $\gamma$-ray halos. For fast diffusion even the ballistic trajectory is taking into account it can not explain the $\gamma$-ray halos neither for a good fit to the SBP distribution nor for a reasonable transfer efficiency. 

Then we investigate the impact of the slow diffusion on cosmic rays propagation. We reanalyze the pulsar scenario to account for the positron spectrum on the Earth in the two-zone diffusion model. A close and old pulsar, like B1055-52, is ideal to make contribution and explain the positron flux.

Furthermore, we study a slow-disk diffusion model considering the populated slow diffusion regions around pulsars in the Galactic disk. The model explains many interesting features in cosmic rays spectra measured by AMS-02. We reexamined the dark matter scenario to account for the positron flux in the slow-disk diffusion model. The dark matter annihilation cross section is much smaller than that in the conventional model and satisfies all the constraints on the annihilation cross section set by $\gamma$-ray and CMB observations. Therefore the dark matter scenario is still feasible for positron excess.


\begin{thebibliography}{99}

\bibitem{HAWC:2017kbo}
A.~U.~Abeysekara \textit{et al.} [HAWC],
Science \textbf{358} (2017) no.6365, 911-914
doi:10.1126/science.aan4880
[arXiv:1711.06223 [astro-ph.HE]].

\bibitem{Yuan:2017ozr}
Q.~Yuan, S.~J.~Lin, K.~Fang and X.~J.~Bi,
Phys. Rev. D \textbf{95} (2017) no.8, 083007
doi:10.1103/PhysRevD.95.083007
[arXiv:1701.06149 [astro-ph.HE]].

\bibitem{LHAASO:2021crt}
F.~Aharonian \textit{et al.} [LHAASO],
Phys. Rev. Lett. \textbf{126} (2021) no.24, 241103
doi:10.1103/PhysRevLett.126.241103
[arXiv:2106.09396 [astro-ph.HE]].

\bibitem{Fang:2019iym}
K.~Fang, X.~J.~Bi and P.~F.~Yin,
Mon. Not. Roy. Astron. Soc. \textbf{488} (2019) no.3, 4074-4080
doi:10.1093/mnras/stz1974
[arXiv:1903.06421 [astro-ph.HE]].

\bibitem{Fang:2018qco}
K.~Fang, X.~J.~Bi, P.~F.~Yin and Q.~Yuan,
Astrophys. J. \textbf{863} (2018) no.1, 30
doi:10.3847/1538-4357/aad092
[arXiv:1803.02640 [astro-ph.HE]].

\bibitem{Evoli:2018aza}
C.~Evoli, T.~Linden and G.~Morlino,
Phys. Rev. D \textbf{98} (2018) no.6, 063017
doi:10.1103/PhysRevD.98.063017
[arXiv:1807.09263 [astro-ph.HE]].

\bibitem{Recchia:2021kty}
S.~Recchia, M.~Di Mauro, F.~A.~Aharonian, L.~Orusa, F.~Donato, S.~Gabici and S.~Manconi,
Phys. Rev. D \textbf{104} (2021) no.12, 123017
doi:10.1103/PhysRevD.104.123017
[arXiv:2106.02275 [astro-ph.HE]].

\bibitem{Bao:2021hey}
L.~Z.~Bao, K.~Fang, X.~J.~Bi and S.~H.~Wang,
Astrophys. J. \textbf{936} (2022) no.2, 183
doi:10.3847/1538-4357/ac8b8a
[arXiv:2107.07395 [astro-ph.HE]].

\bibitem{Fang:2019ayz}
K.~Fang, X.~J.~Bi and P.~F.~Yin,
Astrophys. J. \textbf{884} (2019), 124-128
doi:10.3847/1538-4357/ab3fac
[arXiv:1906.08542 [astro-ph.HE]].

\bibitem{Zhao:2021yzf}
M.~J.~Zhao, K.~Fang and X.~J.~Bi,
Phys. Rev. D \textbf{104} (2021) no.12, 123001
doi:10.1103/PhysRevD.104.123001
[arXiv:2109.04112 [astro-ph.HE]].

\bibitem{Lv:2023alj}
X.~J.~Lv, X.~J.~Bi, K.~Fang, P.~F.~Yin and M.~J.~Zhao,
[arXiv:2307.07114 [astro-ph.HE]].

\bibitem{Liu:2016ngs}
W.~Liu, X.~J.~Bi, S.~J.~Lin and P.~F.~Yin,
Chin. Phys. C \textbf{41} (2017) no.4, 045104
doi:10.1088/1674-1137/41/4/045104
[arXiv:1602.01012 [astro-ph.CO]].

\bibitem{Yin:2013vaa}
P.~F.~Yin, Z.~H.~Yu, Q.~Yuan and X.~J.~Bi,
Phys. Rev. D \textbf{88} (2013) no.2, 023001
doi:10.1103/PhysRevD.88.023001
[arXiv:1304.4128 [astro-ph.HE]].

\bibitem{Posselt:2023dau}
B.~Posselt, G.~G.~Pavlov, O.~Kargaltsev and J.~Hare,
Astrophys. J. \textbf{952} (2023) no.2, 134
doi:10.3847/1538-4357/acd9d1
[arXiv:2306.00185 [astro-ph.HE]].

\bibitem{Xiang:2017jou}
Q.~F.~Xiang, X.~J.~Bi, S.~J.~Lin and P.~F.~Yin,
Phys. Lett. B \textbf{773} (2017), 448-454
doi:10.1016/j.physletb.2017.09.003
[arXiv:1707.09313 [astro-ph.HE]].

\bibitem{Fermi-LAT:2016uux}
A.~Albert \textit{et al.} [Fermi-LAT and DES],
Astrophys. J. \textbf{834} (2017) no.2, 110
doi:10.3847/1538-4357/834/2/110
[arXiv:1611.03184 [astro-ph.HE]].

\bibitem{Fermi-LAT:2012pls}
M.~Ackermann \textit{et al.} [Fermi-LAT],
Astrophys. J. \textbf{761} (2012), 91
doi:10.1088/0004-637X/761/2/91
[arXiv:1205.6474 [astro-ph.CO]].

\bibitem{Slatyer:2015jla}
T.~R.~Slatyer,
Phys. Rev. D \textbf{93} (2016) no.2, 023527
doi:10.1103/PhysRevD.93.023527
[arXiv:1506.03811 [hep-ph]].

\bibitem{Lin:2014vja}
S.~J.~Lin, Q.~Yuan and X.~J.~Bi,
Phys. Rev. D \textbf{91} (2015) no.6, 063508
doi:10.1103/PhysRevD.91.063508
[arXiv:1409.6248 [astro-ph.HE]].


\bibitem{Wang:2018pcc}
B.~Wang, X.~Bi, S.~Lin and P.~Yin,
Sci. China Phys. Mech. Astron. \textbf{61} (2018) no.10, 101004
doi:10.1007/s11433-018-9244-y


\end{thebibliography}
\end{document}